\documentclass[aps,preprint]{revtex4}
\usepackage{epsfig}
\usepackage{color}
\usepackage{subfigure}
\newcommand {\sla}[1]{ #1 \!\!\!/}

\begin{document}

\draft
\title{
{\normalsize \hskip4.2in USTC-ICTS-07-08} \\
{\bf Backward Compton Scattering and QED with Noncommutative Plane
in the Strong Uniform Magnetic Field}}
\author{Wei Huang$^{a}$}
\author{Wang Xu$^{a,b}$}
\author{ Mu-Lin Yan$^{a}\footnote{Corresponding author.
Email address: mlyan@ustc.edu.cn}$}

\affiliation{\centerline{$^a$ Interdisciplinary Center for Theoretical Study,
Department of Modern Physics}
\centerline{University of Science and Technology of China, Hefei, Anhui 230026, China}
\centerline{$^b$ Shanghai Institute of Applied Physics,
Chinese Academy of Sciences, Shanghai, China }}
\begin{abstract}
In the strong uniform magnetic field, the noncommutative plane (NCP)
caused by the lowest Landau level (LLL) effect, and QED with NCP
(QED-NCP) are studied. Being similar to the condensed matter theory
of quantum Hall effect, an effective filling factor $f(B)$ is
introduced to characterize the possibility that the electrons stay
on the LLL. The analytic and numerical results of the differential
cross section for the process of backward Compton scattering in
accelerator with unpolarized or polarized initial photons are
calculated. The existing data of BL38B2 in Spring-8 have been
analyzed roughly and compared with the numerical predictions
primitively. We propose a precise measurement of the differential
cross sections of backward Compton scattering in a strong
perpendicular magnetic field, which may reveal the effects of NCP.

\vskip0.2in



\end{abstract}
\maketitle

\section{Introduction}
The physics related to the lowest Landau level (LLL) and
corresponding spacetime noncommutativity, especially noncommutative
field theory (NCFT), have long been studied with considerable
interest \cite{review}, and appear naturally in fundamental field
theory \cite{NCQED, Miransky} and condensed matter theory
\cite{Susskind, DJ}. Spacetime noncommutativity was proposed by
Heisenberg in the 1930's, in order to introduce an effective
ultraviolet cutoff to control the ultraviolet divergences in quantum
field theory. Peierls applied it to non-relativistic electronic
systems in external magnetic fields, which is the first
phenomenological realization of spacetime noncommutativity, and
Snyder published it with systematic analysis in 1947 \cite{review}.
Recently, noncommutative QED (NCQED) \cite{NCQED} and other NCFTs
have been constructed as limits of string/M theory \cite{review},
and as the LLL approximation of QED or the Nambu-Jona-Lasinio model
in the strong magnetic field \cite{Miransky}. In condensed matter
theory, NCFT, particularly the noncommutative Chern-Simons theory
\cite{Susskind}, provides a better mean field theory description of
the fractional quantum Hall states, which can reproduce the detailed
properties and the correct quantitative features of quasiparticles.
In the present paper, we try to explore the effect of space
noncommutativity caused by the LLL, and the possibility to measure
it by considering backward Compton scattering in the external
magnetic field in accelerator.

Considering a non-relativistic electron in a uniform magnetic field
\cite{LL},
\begin{equation}\label{1}
L={1\over 2}m_e(\dot{x}^2 +\dot{y}^2 + \dot{z}^2) +{e\over
c}(\dot{x}A_x +\dot{y}A_y +\dot{z}A_z)-V(x,z),~
\overrightarrow{A}=(0, 0, -xB)
\end{equation}
or a non-relativistic 2D electronic system in a perpendicular
magnetic field \cite{review},
\begin{equation}\label{2}
L = \sum_{{\mu}=1}^{N_e}{1 \over 2} m_e {\dot{\vec x}}^2_{\mu}
- {i e B \over 2c} {\varepsilon}_{ij} x^i_{\mu} {\dot x}^j_{\mu} + V(\vec x_\mu) \\
+ \sum_{{\mu} < {\nu}} U ( {\vec x}_{\mu} - {\vec x}_{\nu} ),
\end{equation}
the energy eigenvalues of the Landau Levels are:
\begin{equation}\label{3}
E_{n}=\hbar {eB\over m_e c}(n+ {1\over 2}).
\end{equation}
In the limit of the strong magnetic field, the separation between
the Landau levels becomes very large and consequently only the LLL
is relevant. One can neglect the kinetic term, i.e. formally put
$m_e=0$, the resulting Lagrangian is first order in time
derivatives, turning the original coordinate space into an effective
phase space defined by:
\begin{equation}\label{4}
p_z\equiv{\partial L_{LLL}\over \partial \dot{z}}=-{eB\over c}x
~\Rightarrow ~\left[-{eB\over c}x, z\right]=-i\hbar ~\Rightarrow
~\left[ x, z\right]=i{\hbar c\over eB}\equiv i\theta_L ,
\end{equation}
or
\begin{equation}\label{5}
[x^i_{\mu}, x^j_{\nu}] = i{\delta}_{\mu\nu} {\varepsilon}^{ij}
{{\hbar} c \over {eB}} \equiv i{\delta}_{\mu\nu}
{\varepsilon}^{ij}\theta_L .
\end{equation}

The effects of truncation to the LLL are now expressed by
noncommutativity, which is described by $\theta_L= {\hbar c\over
eB}$. It is essential that the equations (\ref{4},\ref{5}) indicate
that in the 3-dimensional space there is a noncommutative plane
(NCP) perpendicular to the strong external magnetic field $B$.

The existence of NCP has been widely used to discuss the quantum
Hall effect and relevant topics in condensed matter physics and
mathematical physics \cite{Susskind, DJ}. In such discussions on the
quantum Hall effect, the noncommutative parameter for NCP is usually
taken to be
\begin{equation}\label{6}
\theta=f\theta_L,
\end{equation}
where $f=f(\nu, B)$ is a function of the filling fraction $\nu$ and
the magnetic field $B$, e.g. $f = \frac{1}{\nu} =
\frac{eB}{2\pi\rho}$ in the noncommutative Chern-Simons theory
\cite{Susskind}, and it could be thought as an effective filling
factor to characterize the possibility that the electrons stay on
the LLL. At $f=0$, no electron stays on the LLL, so that the NCP
caused by the external magnetic field $B$ is absent. For $f\neq 0$,
the NCP exists and must be considered. In this paper $f(B)$ is
treated as a phenomenology parameter.

A nature question arising from the condensed matter physics
discussions mentioned above is whether such sort of NCP discussions
can be extended into the QED dynamics of electron beam in
accelerator, where the electrons are correlative to each other. It
is always a possibility that some electrons stay on the LLL and
$f\neq 0$, and there is no prior reason to ban this extension, hence
the answer should be yes. As a matter of fact \cite{WY}, the
anomalous deviation of (g-2)-factor of muon to the prediction of the
standard model has been attributed to the loop effects of QED with
NCP. That could be thought as a rough estimation of the NCP effects
in QED at loop level. However, the loop level process has some
uncertainties both due to the theoretical treatment errors and the
experimental measurement errors, and a tree level process in the
accelerator experiments could be essential to make it clear. Hence,
we consider the backward Compton scattering process in the strong
magnetic field, e.g. the beamline BL38B2 in Spring-8, to explore
whether the NCP effects exist or not.

The point for revealing the NCP effects caused by the LLL effect in
a process is that the perpendicular external magnetic field $B$
``felt'' by the correlated electrons with {\it non-relativistic}
motion should be very strong. As the backward Compton scattering is
a process that the soft laser photons are backscattered by the high
energy electrons elastically, the motion of the electrons in the
$e\gamma$-mass center frame (CM) is non-relativistic, the Lorentz
factor to the laboratory frame is very large and the magnetic field
``felt'' by the electrons $B=B_{CM}=\gamma B_{Lab}$ becomes very
large even if $B_{Lab}$ is small. For instance, in the mass center
frame of the beamline BL38B2 in Spring-8 with 8GeV electron, 0.01eV
photon and 0.68T magnetic field, the velocity of the electron
$v_{CM}\approx 0.0006\ll 1,\;\gamma\approx 15645.6,\; B_{CM}\approx
10639T$. It well satisfies the precondition, hence the NCP due to
the LLL could be described by a noncommutative quantum theory
constructed in the mass center frame.

The contents of this paper are organized as follows: in Section II,
we construct QED with NCP; in Section III, we derive the
differential cross section of the backward Compton scattering
process in a uniform perpendicular magnetic field; in Section IV, we
produce the numerical results on it by using the data of Spring-8,
and show how a precise measurement of the differential cross section
leads to distinguishing the prediction of QED with NCP from the
prediction of QED without NCP; finally, we briefly discuss the
results.

\section{QED with NCP}
In order to construct the effective Lagrangian describing the
electrons in the external magnetic field, the LLL effect should be
considered. For the electrons stay on the LLL, the effects of
projection on the LLL could be expressed by noncommutativity
(natural units $\hbar=c=1$):
\begin{equation}\label{7} [\hat{x}_\mu,
\hat{x}_\nu]=i\theta_{\mu\nu}=i\theta C_{\mu\nu},~~~
\theta=f\theta_L=f{1\over eB},
\end{equation}
\begin{equation}\label{8}
C_{\mu\nu}=\left( \begin{array}{cccc}
             0 & c_{01} & c_{02} & c_{03} \\
             -c_{01} & 0 & c_{12} & -c_{13} \\
             -c_{02} & -c_{12} & 0 & c_{23} \\
             -c_{03} & c_{13} & -c_{23} & 0
             \end{array} \right).
\end{equation}

The Lagrangian in a noncommutative theory is fully covariant
under observer Lorentz transformations: rotations or boosts of the
observer inertial frame leave the physics unchanged because both the
field operators and $\theta_{\mu\nu}$ transform covariantly \cite{LV}.
In this paper, we calculate in the mass center frame, in
which the motion of the electron is non-relativistic and only
$\theta_{ij}$ are nonzero, and finally boost the results to the
laboratory frame to compare with the experiment. The direction of
the external magnetic field is $\hat{y}$ in the laboratory frame, by
means of the Lorentz transformation, the electron feels an electric
field along $-\hat{x}$ and a magnetic field along $\hat{y}$ in the
mass center frame. The electric field has no influence on the
noncommutativity caused by the LLL \cite{DJ}, so that $c_{0i}=0$.
The magnetic field is along $\hat{y}$ and the NCP takes
$(x,\;z)$-plane, so that $c_{13}=1$ and other $c_{ij}=0$.

Generally \cite{NCQED}, we can implement the noncommutativity of
space into path integral formulation through the Weyl-Moyal
correspondence, and the noncommutative version of a field theory
can be obtained by replacing the product of the fields appearing in
the action by the star product:
\begin{equation}\label{9}
(f*g)(x)=\lim_{\xi,\eta\rightarrow 0} \left[e^{\frac{i}{2}\partial_\xi^\mu
\theta_{\mu\nu} \partial_\eta^\nu} f(x+\xi) g(x+\eta)\right].
\end{equation}

Following the general argument, we argue that the effective
Lagrangian of QED with NCP (QED-NCP) for the electrons with
$f(B)\neq 0$ should be an extension of the Lagrangian of NCQED with
$f(B)$:
\begin{equation}\label{10}
\mathcal{L}=-{1\over 4}F_{\mu\nu}*F^{\mu\nu}+
\overline{\psi}*(i\gamma^\mu D_\mu-m)*\psi,
\end{equation}
with \begin{equation}\label{11} D_\mu=\partial_\mu-ieA_\mu,~~~
F_{\mu\nu}=\partial_\mu A_\nu- \partial_\nu A_\mu - i e[A_\mu,
A_\nu]_*.
\end{equation}

The above Lagrangian is invariant under the noncommutative U(1) transformation:
\begin{eqnarray*}
&& A_\mu \to A'_\mu(x) = U(x)*A_\mu*U(x)^{-1} + i U(x)*\partial_\mu U(x)^{-1} ,\\
&& F_{\mu \nu} \to F_{\mu \nu}' = U(x)*F_{\mu \nu}*U(x)^{-1} ,\\
&& \Psi(x) \to \Psi'(x) = U(x)*\Psi(x) ,\\
&& U(x) = \exp *(i\lambda(x)) \equiv 1+
i\lambda(x)-\frac{1}{2}\lambda(x)*\lambda(x) + o(\theta^2) .
\end{eqnarray*}

Note that when $f(B)\rightarrow 0$, the Lagrangian of QED-NCP goes
back to the ordinary QED Lagrangian. When $f(B)\ll 1$, the deviation
of QED-NCP from QED can be calculated in perturbation, but no vacuum
phase transition takes place. When $B$ is extremely large (e.g.
$\sim 10^{9}T$), $f(B)\sim 1$ and the dynamical symmetry breaking
may occur \cite{Miransky}.

\section{Backward Compton scattering}
\begin{figure}[!htbp]
\subfigure[$i e\gamma^{\mu}\exp({i p_1\theta p_2/2})$]
{\label{fr:mini:subfig:a}
\begin{minipage}[b]{0.4\textwidth}\centering
\includegraphics*[width=0.5\textwidth]{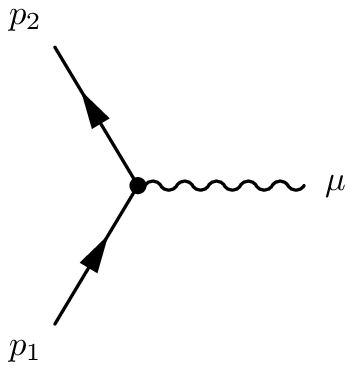}\end{minipage}}%
\hspace{0.1\textwidth} \subfigure[$2 e\sin(k_1\theta
k_2/2)((k_1-k_2)^\rho g^{\mu\nu}+(k_2-k_3)^\nu
g^{\rho\mu}+(k_3-k_1)^\mu g^{\nu\rho})$] {\label{fr:mini:subfig:b}
\begin{minipage}[b]{0.4\textwidth}\centering
\includegraphics*[width=0.5\textwidth]{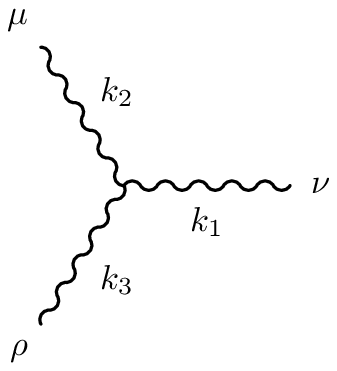}\end{minipage}}
\caption{Feynman rules} \label{fr:mini:subfig}
\end{figure}

From the Lagrangian Eq.(\ref{10}), the Feynman rules of QED-NCP can
be obtained. The propagators of electron and photon remain
unchanged, the vertices in QED-NCP (see Fig.\ref{fr:mini:subfig})
pick up additional kinematic phases from the Fourier transformation
of new interactions. When the inverse Compton scattering by external
electromagnetic fields or the synchrotron radiation is investigated,
the $A_\mu$ in the Lagrangian of QED-NCP should be replaced by $
A_\mu +A_\mu^{external}$. In this paper we do not study those
processes, but only interest in the Compton scattering process,
hence the $A_\mu^{external}$ and the four photon vertex are
neglected.

Similar to the existed calculations of Compton scattering in
NCQED \cite{Compton}, the Feynman diagrams, kinematics and the
differential scattering cross section for the backward Compton
scattering process in QED-NCP are as follows:

\begin{enumerate}

\item The Feynman diagrams of $e\gamma$-Compton scattering
in QED-NCP are shown in Fig.\ref{fg:mini:subfig}. $\mathcal{A}_i$
with $i=1,2,3$ denote the amplitudes of corresponding diagrams.
Compared with that in QED, there is an additional diagram
$\mathcal{A}_3$ (see Fig.\ref{fg:mini:subfig:c}).
\begin{figure}[!htbp]
\subfigure[$\mathcal{A}_1$]{ \label{fg:mini:subfig:a}
\begin{minipage}[b]{0.3\textwidth}\centering
\includegraphics*[width=1.5in]{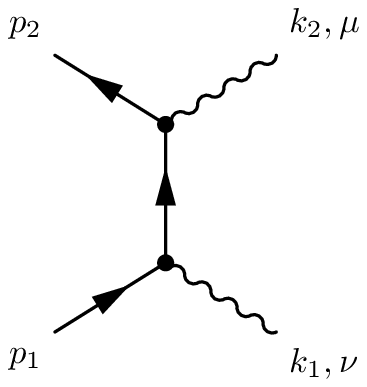}\end{minipage}}%
\subfigure[$\mathcal{A}_2$]{ \label{fg:mini:subfig:b}
\begin{minipage}[b]{0.3\textwidth}\centering
\includegraphics*[width=1.5in]{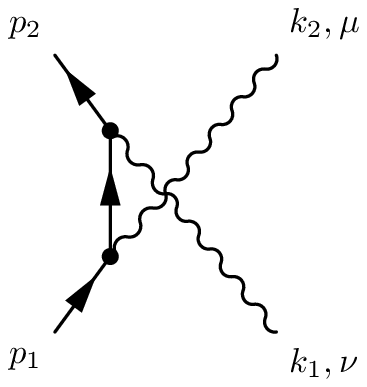}\end{minipage}}%
\subfigure[$\mathcal{A}_3$] { \label{fg:mini:subfig:c}
\begin{minipage}[b]{0.3\textwidth}\centering
\includegraphics*[width=1.5in]{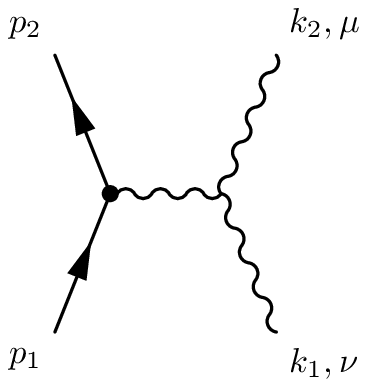}\end{minipage}}
\caption{Feynman diagrams} \label{fg:mini:subfig}
\end{figure}

\item Kinematics (see Fig.\ref{kin:mini:subfig}):
\begin{figure}[!htbp]
\subfigure[The laboratory frame]{\label{kin:mini:subfig:a}
\begin{minipage}[b]{0.5\textwidth}\centering
\includegraphics*[width=2.0in]{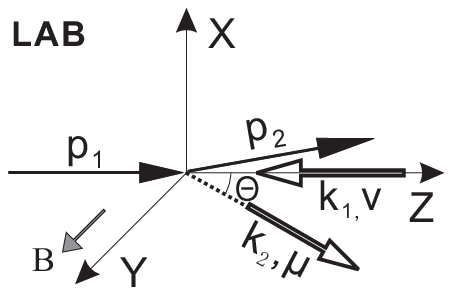}\end{minipage}}%
\subfigure[The mass center frame]{\label{kin:mini:subfig:b}
\begin{minipage}[b]{0.5\textwidth}\centering
\includegraphics*[width=2.0in]{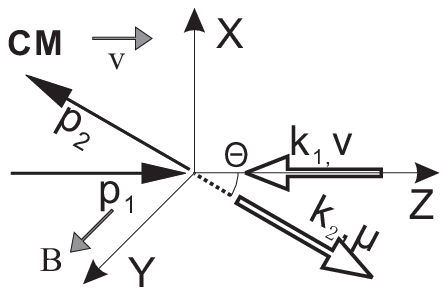}\end{minipage}}
\caption{Kinematics} \label{kin:mini:subfig}
\end{figure}

i) The energies and momenta in the mass center frame:
\begin{eqnarray*}
 && s=(p_1+k_1)^2,~~t=(p_1-p_2)^2,~~u=(p_1-k_2)^2,\\
 && p_1=(\frac{s+m^2}{2\sqrt{s}}, 0, 0, \frac{s-m^2}{2\sqrt{s}} ),~
 k_1=\frac{s-m^2}{2\sqrt{s}}(1, 0, 0, -1 ),\\
 && p_2=\frac{s-m^2}{2\sqrt{s}}(\frac{s+m^2}{s-m^2},
 -\sin\vartheta\cos\phi, -\sin\vartheta\sin\phi, -\cos\vartheta ),\\
 && k_2=\frac{s-m^2}{2\sqrt{s}}(1,\sin\vartheta\cos\phi,
 \sin\vartheta\sin\phi, \cos\vartheta).
\end{eqnarray*}

ii) Polarization: We are interested in the process with polarized
initial electrons, unpolarized or $\alpha$-polarized initial photons
($\alpha$ is the angle between the magnetic field and the initial
photon polarization), unpolarized final electrons and unpolarized
final photons. So the following notations and formulas will be
useful for our goal:
\begin{eqnarray*}
 && {\rm 1)~initial ~electron :~} u_{-1/2}(p_1)\bar{u}_{-1/2}(p_1)
 \rightarrow\rho = \frac{1}{2}(\sla p_1+m)(1-\gamma^5(-1)\gamma^2)\\
 && {\rm 2)~final ~electron :~} \sum_i u_i(p_2)\bar{u}_i(p_2)
 \rightarrow\rho'=\sla p_2+m\\
 && {\rm 3)~initial ~photon :~}
 \frac{1}{2}\sum_i \epsilon^T_{i\mu}(k_1)\epsilon^{T*}_{i\mu'}(k_1)
 \rightarrow\xi_{\mu\mu'}\\
 && or ~{\rm \alpha -polarized :~}
 \epsilon^T_{\alpha\mu}(k_1)\epsilon^{T*}_{\alpha\mu'}(k_1)
 \rightarrow  \xi_{\mu\mu'},~ \epsilon^T_{\alpha\mu}=
 ( 0 , \sin\alpha , \cos\alpha , 0)\\
 && {\rm 4)~final ~photon:~}
 \sum_i\epsilon^T_{i\nu}(k_2) \epsilon^{T*}_{i\nu'}(k_2)
 \rightarrow\xi'_{\nu\nu'}
\end{eqnarray*}

\item The differential cross section for the backward Compton
scattering in QED-NCP is
\begin{equation}\label{12} \frac{d\sigma}{d\phi d\cos\vartheta}
=\frac{e^4}{64\pi^2 s}
\xi_{\mu\mu'}\xi'_{\nu\nu'}
Tr(\rho'\mathcal{A}^{\mu\nu}\rho\bar{\mathcal{A}}^{\nu'\mu'}),
\end{equation}
where $\mathcal{A}^{\mu\nu}=\mathcal{A}^{\mu\nu}_1+
\mathcal{A}^{\mu\nu}_2 + \mathcal{A}^{\mu\nu}_3$ and
$\mathcal{A}^{\mu\nu}_i$, $\bar{\mathcal{A}}^{\nu'\mu'}_i~(i=1,2,3)$
are:
\begin{eqnarray*}
 \mathcal{A}^{\mu\nu}_1 &=& (-1)e^{ip_1\theta p_2/2} e^{ik_1\theta p_2/2}
 \gamma^\mu \frac{\sla p_1+\sla k_1+m}{(p_1+k_1)^2 - m^2}\gamma^\nu\\
 \mathcal{A}^{\mu\nu}_2 &=& (-1)e^{ip_1\theta p_2/2}e^{-ik_1\theta p_2/2}
 \gamma^\nu\frac{\sla p_1-\sla k_2+m}{(p_1-k_2)^2 - m^2}\gamma^\mu\\
 \mathcal{A}^{\mu\nu}_3 &=& (-i)e^{ip_1\theta p_2/2} 2\sin(k_1\theta k_2/2)
 \gamma^\sigma [g_{\rho\sigma}/(k_1-k_2)^2]\\
 &&[(k_1+k_2)^\rho g^{\mu\nu}+(k_1-2k_2)^\nu g^{\rho\mu}+(k_2-2k_1)^\mu g^{\nu\rho}]\\
 \bar{\mathcal{A}}^{\nu'\mu'}_1 &=& (-1)e^{-ip_1\theta p_2/2}e^{-ik_1\theta p_2/2}
 \gamma^{\nu'} \frac{\sla p_1+\sla k_1+m}{(p_1+k_1)^2-m^2}\gamma^{\mu'}\\
 \bar{\mathcal{A}}^{\nu'\mu'}_2 &=& (-1)e^{-ip_1\theta p_2/2}e^{+ik_1\theta p_2/2}
 \gamma^{\mu'} \frac{\sla p_1-\sla k_2+m}{(p_1-k_2)^2-m^2}\gamma^{\nu'}\\
 \bar{\mathcal{A}}^{\nu'\mu'}_3 &=& (i)e^{-ip_1\theta p_2/2} 2\sin(k_1\theta k_2/2)
 \gamma^{\sigma'} [g_{\rho'\sigma'}/(k_1-k_2)^2]\\
 &&[(k_1+k_2)^{\rho'} g^{\mu'\nu'}+(k_1-2k_2)^{\nu'} g^{\rho'\mu'}
 +(k_2-2k_1)^{\mu'} g^{\nu'\rho'}]
\end{eqnarray*}
We define the phase factor $\Delta\equiv \frac{k_1\theta
p_2}{2}=-\frac{k_1\theta k_2}{2} = \frac{f(s-m^2)^2}{8B e s}
\sin\vartheta\cos\phi$ (notation $k\theta p \equiv k^\mu
\theta_{\mu\nu}p^\nu$), and then the differential cross sections of
the backward Compton scattering with polarized initial electrons,
unpolarized initial photons, unpolarized final electrons and
unpolarized final photons in QED-NCP are:
\begin{eqnarray}\label{13}
\nonumber \frac{d\sigma}{d\phi d\cos\vartheta}& = &
\frac{e^4}{32\pi^2 s} \left((s-m^2)^2+(u-m^2)^2 -
\frac{4m^2 t(m^4-s u)} {(s-m^2)(u-m^2)}\right)\\
& \times & \left(- \frac{1}{(s-m^2)(u-m^2)}+\frac{4\sin^2\Delta}{t^2}\right).
\end{eqnarray}
Note that it's $f(B)$ dependent and goes back to that in QED as
$f(B)\rightarrow 0$, and coincides with that in NCQED \cite{Compton}
as $m\rightarrow 0$. Similarly, for the processes with any
polarization, the differential cross sections could be calculated,
some numerical results are as follows.
\end{enumerate}

\section{Numerical Results}
\begin{figure}[!htbp]
\begin{center}
\includegraphics*{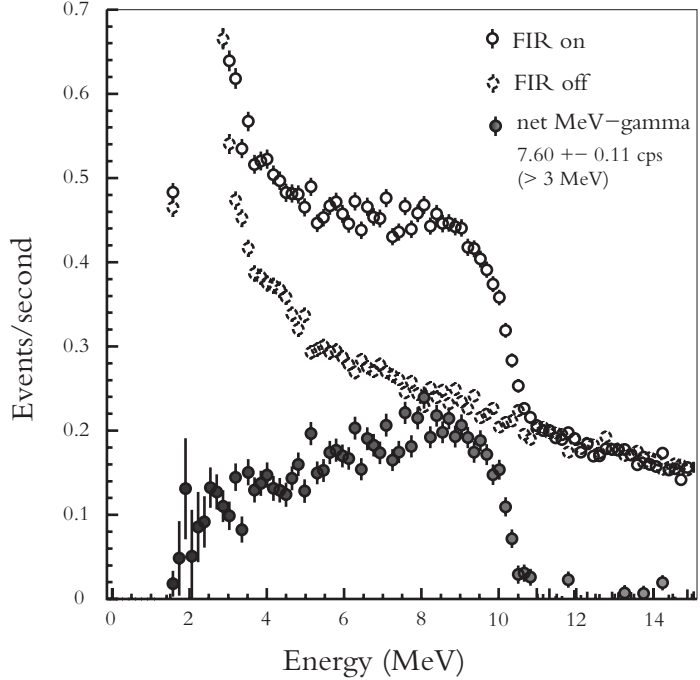}
\caption{Spring-8 data for $e\gamma\rightarrow e'\gamma'$
\cite{spring8}. The $\gamma$-ray spectrum from the backward Compton
scattering process has been deduced after the subtraction of the
``FIR laser off'' spectrum from the ``FIR laser on'' spectrum. They
are shown by the solid circles and proportional to
${d\sigma(E_\gamma) \over dE_\gamma}$.}\label{spring8}
\end{center}
\end{figure}

In this section, the data of BL38B2 in Spring-8 will be used to
discuss the QED-NCP predictions of backward Compton scattering
numerically. The accelerator diagnosis beamline BL38B2 in Spring-8
has a bending magnet light source, $10 MeV \gamma$-ray photons are
produced in the magnetic field by the backward Compton scattering of
far-infrared (FIR) laser photons. The energy of electron in the
storage ring is $8GeV$, the perimeter of the ring is $1436m$, the
wavelength of FIR laser photon is $119\mu m$ and the magnetic field
is $0.68T$. Then, in the mass center frame, the Lorentz factor
$\gamma \approx 15645.6$, the magnetic field is $2\times 10^6
eV^2\approx 10639T$ (hence the LLL effect is relevant), $\theta_L$
is $1.6\times10^{-6}eV^{-2}\approx (2.5 {\rm \AA})^2$ and the phase
factor becomes $\Delta \approx 0.0844 f \sin\vartheta\cos\phi$.
Substituting all of these into Eq.(\ref{12}), the realistic
calculations are doable. Fig.\ref{spring8} shows a measurement of
the differential cross section to final photon energy of the
backward Compton scattering in Spring-8, in order to compare with
it, the $\phi$-integrated energy dependence of the differential
cross section is calculated.

Suppose the initial photon is unpolarized, from Fig.\ref{spring8},
we can roughly see:
\begin{equation}\label{14}
\mathcal{R}|_{expt} =
\frac{d\sigma(5MeV)/dE_\gamma}{d\sigma(9MeV)/dE_\gamma}|_{expt}
\approx \frac{0.15}{0.22} \approx0.68 .
\end{equation}

\begin{figure}[!htbp]
\subfigure[Unpolarized initial photon]
{\label{ncp:mini:subfig:a}\begin{minipage}[b]{0.5\textwidth}\centering
\includegraphics*[width=3.0in]{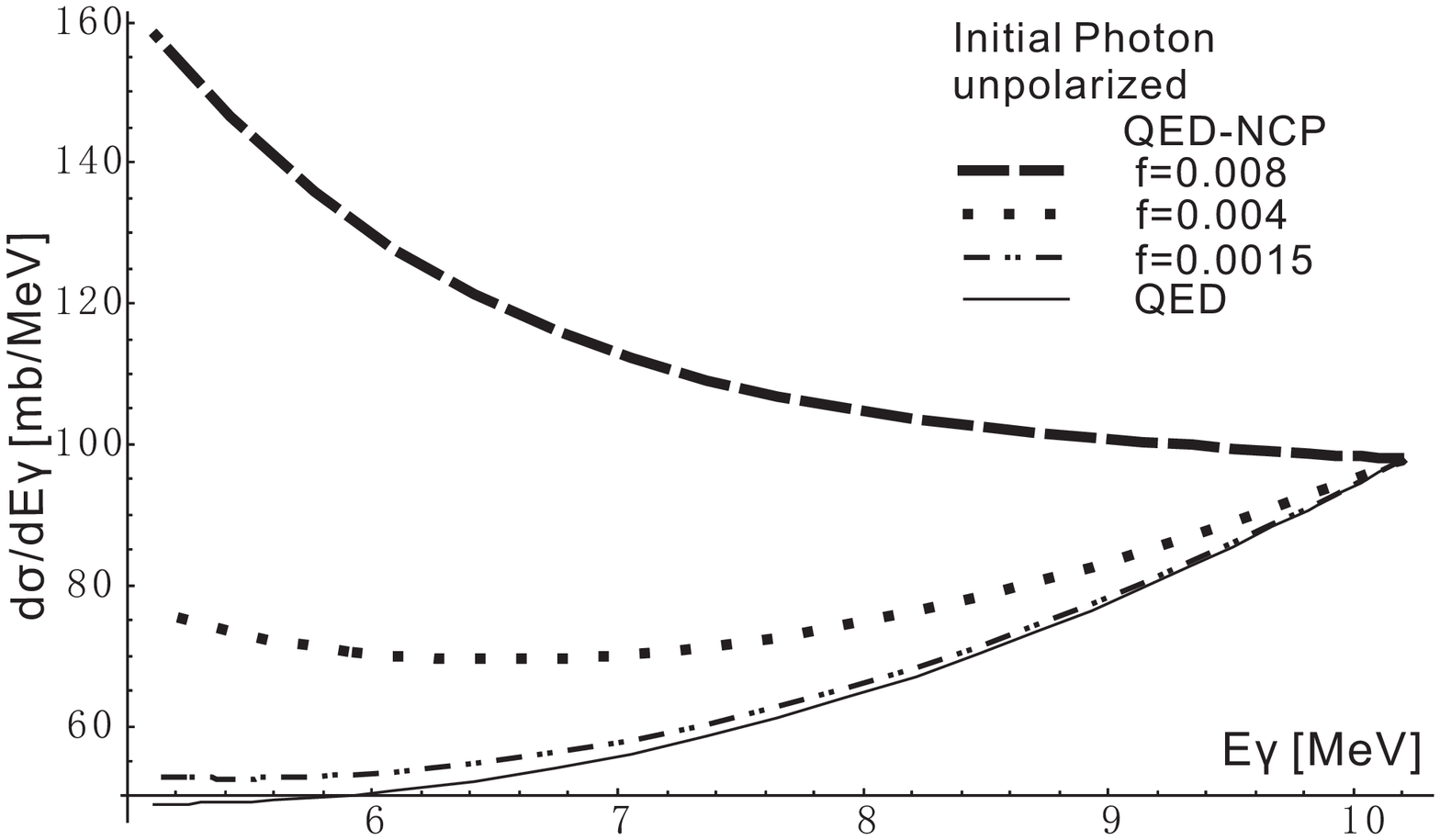}\end{minipage}}%
\subfigure[$\hat{x},\hat{y}$-polarized initial photon]
{\label{ncp:mini:subfig:b}\begin{minipage}[b]{0.5\textwidth}\centering
\includegraphics*[width=3.0in]{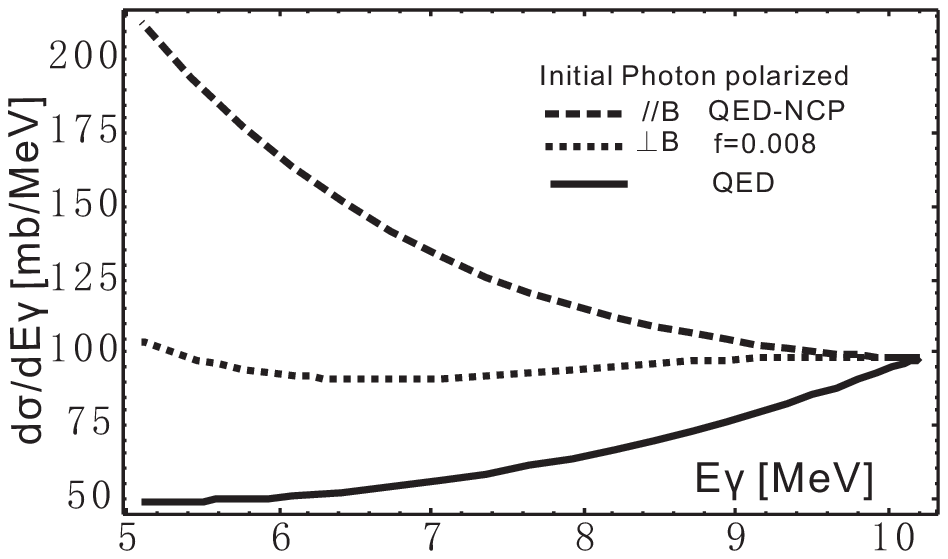}\end{minipage}}
\caption{Energy dependence of the differential cross section.}
\label{ncp:mini:subfig}
\end{figure}

However, we find out that $\mathcal{R} |_{expt}$ is significantly
larger than the QED prediction (Fig.\ref{ncp:mini:subfig:a}):
\begin{equation}\label{15}
\mathcal{R}|_{QED} = \frac{d\sigma(5MeV)/dE_\gamma}{d\sigma(9MeV)
/dE_\gamma}|_{QED} \approx\frac{48.87}{77.43} \approx 0.63 .
\end{equation}

A natural interpretation to this deviation is that the possibility
that the electrons stay on LLL is nonzero, and there is a NCP in the
external magnetic field, which hasn't been taken into account in
QED. By means of QED-NCP, and adjusting the effective filling factor
$f(B)$, a suitable $\mathcal{R}|_{QED-NCP}$ consistent with
$\mathcal{R}|_{expt}$ can be obtained. The corresponding prediction
with $f(B)=0.0015$ is shown in Fig.\ref{ncp:mini:subfig:a}:
\begin{equation}\label{16}
\mathcal{R}|_{QED-NCP} = \frac{d\sigma(5MeV)/dE_\gamma}
{d\sigma(9MeV) /dE_\gamma}|_{QED-NCP} \approx \frac{52.88}{78.24}
\approx0.68 .\end{equation}

However, photon polarization, detector inefficiency and radiation
corrections due to mirror and windows will all affect the shapes of
experimental data, the uncertainties of current experimental data
are too large to separate two calculations. It is still too early to
decide the existence of the NCP effects. A further precise
measurement is needed.

Theoretically, for a 2D electronic system, $f=\frac{eB}{2\pi\rho}$
proved in \cite{Susskind} can be used. The electron beam of BL38B2
in Spring-8, whose charge is around $1.44nC$, length is $13ps$,
horizontal size is $114\mu m$ and vertical size is $14\mu m$ ($\ll$
horizontal size), is a near 2D electronic system. Hence, a rough
prediction of $f(B)$ could be calculated:
\begin{eqnarray}
\label{17}\rho& \approx& \frac{1.44 nC/(1.6\times 10^{-19}
C)}{13ps\times
(3\times 10^8 m/s)\times 114 \mu m}\approx 2\times 10^{12} cm^{-2} ,\\
\label{18}f& \approx& \frac{0.68 T\times(2\times 0.511 MeV \times
5.788\times 10^{-11}MeV/T)}{2\pi \times 2\times 10^{12}
cm^{-2}\times (197.3 MeV\times fm)^2}\approx 0.008 .
\end{eqnarray}

With a typical $f(B)=0.008$, we further consider experiments with
polarized initial photon. The initial laser photons move along the
direction of $\hat{z}$ and their polarization is taken either
parallel or perpendicular to the magnetic field direction of
$\hat{y}$. As shown in Fig.\ref{ncp:mini:subfig:b}, the energy
dependence of the differential cross sections with the
$\hat{x}$-polarized ($\perp \hskip-0.06in B$) and the
$\hat{y}$-polarized ($\parallel \hskip-0.06in B$) initial photons
are the same in QED, and different in QED-NCP. This strongly
suggests that a precise backward Compton scattering experiment in
Spring-8 with differently polarized initial photons is most
favorable for testing the NCP effects. The experiment with different
initial photon polarization is practicable to reveal the NCP
effects, because the subtraction of the $\perp \hskip-0.06in
B$-polarized spectrum from the $\parallel \hskip-0.06in B$-polarized
spectrum can reduce the experimental uncertainties.

Furthermore, we consider the total cross sections (barn) by
integrating $E_\gamma$ from $5.1 MeV$ to $10.2 MeV$ (or integrating
$\vartheta$ from $0$ to $\pi /2$):
\begin{eqnarray*}
\sigma_{QED} = \sigma_{QED}^{\perp B} = \sigma_{QED}^{\parallel B}
\approx 0.586936,& & \sigma_{QED-NCP}^{\perp B}\approx 0.586936 + 2828.44 f^2,\\
\sigma_{QED-NCP}\approx 0.586936 + 4384.20 f^2, & &
\sigma_{QED-NCP}^{\parallel B}\approx 0.586936 + 5939.96 f^2,\\
\sigma_{QED-NCP} -\sigma_{QED}\approx 7469.64 f^2 \sigma_{QED}, & &
\sigma_{QED-NCP}^{\parallel B} - \sigma_{QED-NCP}^{\perp B} \approx
5301.30 f^2 \sigma_{QED}.
\end{eqnarray*}
From above we can see that the difference between the total cross
sections of QED and QED-NCP is proportional to $f^2$, and the
difference between the total cross sections with the $\perp
\hskip-0.06in B$-polarized initial photons and with the $\parallel
\hskip-0.06in B$-polarized ones is proportional to $f^2$, too, hence
$f(B)$ characterizing the NCP effects could also be determined in
the $e\gamma$-total cross section measurements.

\section{Summary and discussion}
In this paper, the NCP caused by the LLL effect in the strong
uniform perpendicular magnetic field, and QED with NCP are studied.
For the process of backward Compton scattering in the magnetic field
of the storage ring magnet in accelerator, the amplitudes and the
differential cross sections in QED-NCP are calculated. Numerical
predictions of the energy dependence of the differential cross
sections in QED-NCP and in QED are calculated with the parameters of
BL38B2 in Spring-8, and compared with the existing data of BL38B2.
It indicates that a precise measurement of the energy dependence of
the differential cross sections of backward Compton scattering with
polarized photon in a strong perpendicular magnetic field would be
practicable to distinguish the prediction of QED with NCP from the
prediction of QED without NCP and may reveal the effects of NCP.
Such an experiment is expected.

Being similar to the noncommutative Chern-Simons theory of the
fractional quantum Hall effect, an effective filling factor $f(B)$
is introduced to characterize the possibility that the electrons
stay on the LLL. In this paper $f(B)$ is treated as a phenomenology
parameter and expected to be determined experimentally. A further
task is to estimate it theoretically. In Section IV, we present a
rough estimation of it for BL38B2 in Spring-8. It seems to be
reasonable for near 2D correlated electrons with non-relativistic
motion in the external magnetic field, and supports the NCP
discussion of backward Compton scattering in accelerator. However,
the equation (\ref{17}) is a rough approximative estimation of the
2D electron density under the assumption that the electron beam is
evenly distributed in a finite 2D rectangle, i.e.,
$\rho(x,z)|_{(x,z) \in \;rectangle} = constant$. In a real beamline,
however, the 2D density should be electron-distribution dependent,
e.g., with a Gaussian distribution, we may need to correct the
density $\rho$ in Eq.(\ref{17}) with a factor $\alpha$, i.e., $\rho
\rightarrow \alpha\rho$, where $\alpha=1/2\pi$ or $1/4\pi$. In this
case, the numerical results of $d\sigma/dE_\gamma$ in
Fig.\ref{ncp:mini:subfig} will receive a correction from $\alpha$.
We argue that this correction would not lead to the change of the
basic scenario of $d\sigma/dE_\gamma$ due to QED-NCP. The
discussions in Section IV are instructive, but a more sound
theoretical study on $f(B)$ for the electrons in accelerator is
still wanted, and a detailed discussion on the effects of NCP
remains to be further explored.

\begin{acknowledgments}
We would like to acknowledge Prof. Mamoru Fujiwara for discussion.
One of us (MLY) would like to thank Prof. Yong-Shi Wu for helpful
discussions on the quantum Hall effects. This work is supported by
the National Natural Science Foundation of China under Grant Numbers
90403021, PhD Program Funds of the Education Ministry of China,
Pujiang Talent Project of the Shanghai Science and Technology
Committee under Grant Numbers 06PJ14114, and Hundred Talents Project
of Shanghai Institute of Applied Physics.
\end{acknowledgments}

%
%
%
%
%
\end{document}